\documentclass[12]{article}
\usepackage{fleqn}
\usepackage{graphicx}
\begin{document}
\Large
\begin{center}
{\bf Finite Geometry Behind the Harvey-Chryssanthacopoulos Four-Qubit Magic Rectangle}
\end{center}
\vspace*{.2cm}
\large
\begin{center}
Metod Saniga$^{1}$ and Michel Planat$^{2}$
\end{center}
\vspace*{-.4cm} \normalsize
\begin{center}
$^{1}$Astronomical Institute, Slovak Academy of Sciences\\
SK-05960 Tatransk\' a Lomnica\\ Slovak Republic\\
(msaniga@astro.sk)

\vspace*{.1cm} and

\vspace*{.1cm}
$^{2}$Institut FEMTO-ST/CNRS,
32 Avenue de l'Observatoire\\
F-25044 Besan\c con, France\\
(michel.planat@femto-st.fr)

\end{center}

\vspace*{.0cm} \noindent \hrulefill

\vspace*{.1cm} \noindent {\bf Abstract}

\noindent A ``magic rectangle" of eleven observables of four qubits, employed by Harvey and Chryssanthacopoulos (2008) to prove the Bell-Kochen-Specker theorem in a 16-dimensional Hilbert space, is given a neat finite-geometrical reinterpretation in terms of the structure of the symplectic polar space $W(7, 2)$ of the real four-qubit Pauli group. Each of the four sets of observables of cardinality five represents an elliptic quadric in the three-dimensional projective space of order two (PG$(3, 2)$) it spans, whereas the remaining set of cardinality four corresponds to an affine plane of order two. The four ambient PG$(3, 2)$s of the quadrics intersect pairwise in a line, the resulting six lines meeting in a point. Projecting the whole configuration from this distinguished point (observable) one gets another, complementary ``magic rectangle" of the same qualitative structure.
\\ \\
%{\bf MSC Codes:} 51Exx, 81R99\\
%{\bf PACS Numbers:} 02.10.Ox, 02.40.Dr, 03.65.Ca\\

\vspace*{-.1cm} \noindent \hrulefill

\vspace*{.3cm}
%\large

\section{Introduction}
There exist several ingenious proofs of the famous Bell-Kochen-Specker (BKS) theorem that involve ``magic" configurations of $N$-qubit observables of low ranks. For the $N=2$ case such a configuration is known as the Mermin(-Peres) magic square \cite{mer}, for $N=3$ as the Mermin pentagram \cite{kp,ar} and for $N=4$ as a ``magic rectangle" \cite{hch}. An interesting fact is that by employing the finite symplectic polar space $W(2N -1,2)$ of the generalized Pauli group of $N$-qubits \cite{sp,hos,thas,pla}, the configurations of the first two cases were found to correspond to some distinguished finite geometries. The $N=2$ configuration has several isomorphic finite-geometrical descriptions, namely: a special kind of geometric hyperplane of the symplectic polar space $W(3,2)$ \cite{ps}, a hyperbolic quadric in PG$(3, 2)$ \cite{hos} or, finally, a projective line over the direct product of two smallest Galois fields, $P_{1}(GF(2) \times GF(2))$ \cite{spp}.  The $N=3$ configuration is isomorphic to an ovoid / elliptic quadric of PG$(3,2)$ \cite{sl}. In the present note we shall show that also the Harvey-Chryssanthacopoulos ``magic rectangle" of four-qubit observables \cite{hch} is, as envisaged, underlaid by a remarkable subgeometry of the corresponding finite  symplectic polar space $W(7,2)$ of four qubits \cite{sp,hos,slp}.  This subgeometry, loosely speaking, consists of four concurrent elliptic quadrics situated in four different PG$(3, 2)$s of $W(7, 2)$ that are ``touched," in a particular way, by an affine plane of order two.

The BKS theorem is a significant, but rather subtle, topic in the foundations of quantum mechanics. In one of its formulations \cite{bell,ks}, the theorem asserts that there exist finite sets of projection operators  such that it is impossible to attribute to each one of the operators a bit value, `true' or `false,' subject to the following two constraints: (i) two orthogonal projection operators cannot both be true and (ii) if a subset of orthogonal projection operators is complete, one of these operators must be true. Importance of this theorem for quantum physics lies with the fact that all standard assumptions of local realism and non-contextuality lead to logical contradictions. Hence, any geometrically-oriented insight into its nature is of great interest.

\section{``Magic Rectangle" of Harvey and Chryssanthacopoulos}
To furnish a proof of the BKS theorem in 16 dimensions, Harvey and Chryssanthacopoulos \cite{hch} utilized, arranged in a form of a rectangular array (see also \cite{pla2} for a different rendering),  the following five sets of mutually commuting four-qubit
observables:
\begin{eqnarray*}
& S_1 = \{ZIII, IXII, IIZI, IIIX, ZXZX\},   \\
& S_2 = \{ZIII, IXII, IIXI, IIIZ, ZXXZ\}, \\
& S_3 = \{XIII, IXII, IIZI, IIIZ, XXZZ\}, \\
& S_4 = \{XIII, IXII, IIXI, IIIX, XXXX\}, \\
& S_5 = \{ZXZX, ZXXZ, XXZZ, XXXX\};
\end{eqnarray*}
here
\begin{eqnarray*}
I = \left(
\begin{array}{cc}
1 & 0 \\
0 & 1 \\
\end{array}
\right),~
X = \left(
\begin{array}{cc}
0 & 1 \\
1 & 0 \\
\end{array}
\right),~
Z = \left(
\begin{array}{cc}
1 & 0 \\
0 & -1 \\
\end{array}
\right),
\end{eqnarray*}
and $ABCD$ is a shorthand for the tensor product $A \otimes B \otimes C \otimes D$.
It can readily be checked that the (ordinary matrix) product of observables in any of the first four sets is $+IIII$, whilst that in the last set is $-IIII$. As each observable has eigenvalues of $\pm 1$ and, with the exception of $IXII$, belongs to two different sets (contexts), the above property makes it impossible to assign an eigenvalue to each observable in such a way that the eigenvalues obey  the same multiplication rules as the observables --- such a contradiction providing a proof of the BKS theorem.

Before we embark on geometric considerations it is worth pointing out that whereas in both the $N=2$ (Mermin's square) and $N=3$ (Mermin's pentagram) cases all the sets employed have the same cardinality (three, respectively, four) and each observable belongs to exactly two contexts \cite{mer,kp,ar}, neither of these two properties is met in the present, more involved $N=4$ case. With the help of relevant finite geometry we shall not only discover that sets of different cardinality stem here from {\it qualitatively} different geometric configurations, but also get a clear understanding of the role played by the ``exceptional" observable $IXII$, shared by all $S_i$, $i=1,\ldots, 4$, yet missing in $S_5$.

\section{Geometry of the ``Magic Rectangle"}
To find the finite-geometrical underpinning of the five sets listed in the previous section, we follow the same strategy as in the $N=3$ case \cite{sl}. First, we treat each of the five sets
as a subset of the corresponding (real) four-qubit Pauli group and, employing the $W(7,2)$-geometry of this group \cite{sp,hos,thas,slp}, we look for the totally isotropic subspace of the ambient seven-dimensional projective space PG$(7, 2)$ a given set spans. Next, we figure out a subgeometry a given set corresponds to in the associated subspace. Finally, we analyze how such subspaces are related to each other to deepen our understanding of the revealed geometry.

As per step one, we readily recognize that in each $S_i$, $i= 1, 2, 3, 4$, the ${5 \choose 2} = 10$ products of pairs of observables are all distinct and different from the five observables of the set in question; these $5 + 10 = 15$ mutually commuting observables comprise a $PG(3,2)$, which is {\it maximal} totally isotropic subspace in $W(7, 2)$ \cite{sp,hos}. Explicitly, the set of observables that correspond to the PG$(3,2)_i$ spanned by $S_i$ is:
\begin{eqnarray*}
& {\rm PG}(3,2)_1 =  \{ZIII, IXII, IIZI, IIIX, ZXZX; ZXII, ZIZI, ZIIX, \\
&~~~~~~~~~~~~~~~~ IXZX, IXZI, IXIX, ZIZX, IIZX, ZXIX, ZXZI\},   \\
& {\rm PG}(3,2)_2 = \{ZIII, IXII, IIXI, IIIZ, ZXXZ; ZXII, ZIXI, ZIIZ, \\
&~~~~~~~~~~~~~~~~ IXXZ, IXXI, IXIZ, ZIXZ, IIXZ, ZXIZ, ZXXI\}, \\
& {\rm PG}(3,2)_3 = \{XIII, IXII, IIZI, IIIZ, XXZZ; XXII, XIZI, XIIZ, \\
&~~~~~~~~~~~~~~~~ IXZZ, IXZI, IXIZ, XIZZ, IIZZ, XXIZ, XXZI\}, \\
& {\rm PG}(3,2)_4 = \{XIII, IXII, IIXI, IIIX, XXXX; XXII, XIXI, XIIX,\\
&~~~~~~~~~~~~~~~~ IXXX, IXXI, IXIX, XIXX, IIXX, XXIX, XXXI\}.
\end{eqnarray*}
The situation with $S_5$ is different. Multiplying pairs of these operators one will find that ${4 \choose 2} = 6$ products yield only {\it three} distinct values,\footnote{When speaking in geometrical terms, the sign of an observable is irrelevant, as both $+ABCD$ and $-ABCD$ represents one and the same point of $W(7,2)$ \cite{sp,hos}.} namely $IIYY$, $YIIY$ and $YIYI$; here $Y \equiv XZ$. Hence, the four elements of $S_5$ do not span a PG$(3,2)$, but only a PG$(2,2)$ (the famous Fano plane):
\begin{eqnarray*}
& {\rm PG}(2,2) = \{ZXZX, ZXXZ, XXZZ, XXXX; IIYY, YIIY, YIYI \}.
\end{eqnarray*}

Concerning step two, we shall first address the $S_5$ case. Here we note that since the product  $(IIYY).(YIIY).(YIYI) = - IIII$, the corresponding three points lie on a line in the PG$(2,2)$ (see \cite{slp} for all the essential technicalities of the structure of $W(7,2)$ cast into the group-theoretical setting); hence, the four points of $S_5$ form nothing but an {\it affine plane of order two}. The situation with $S_i$, $i$ running from 1 to 4, is a bit more involved. We have already seen that in any of these sets, the product of any two observables falls off the set.
It can be verified that the same holds with the products of triples of observables. The former property means that no three points of a given set lie on a line, whilst the latter one tells us that
no four points lie in the same plane. A set of five points of PG$(3,2)$ having these properties is another well-known object of finite geometry, namely an {\it elliptic quadric}  (see, e.\,g., \cite{hir,bat}). We thus see that the two kinds of sets involved in the Harvey-Chryssanthacopoulos proof of the BKS theorem are not only related to dimensionally-different ambient totally isotropic subpaces of $W(7,2)$, but they also differ in their intrinsic geometry, an affine plane of order two versus an elliptic quadric of PG$(3,2)$ --- as diagramatically illustrated in Figure 1, {\it left}.

\begin{figure}[t]
\centerline{\includegraphics[width=7.3cm,clip=]{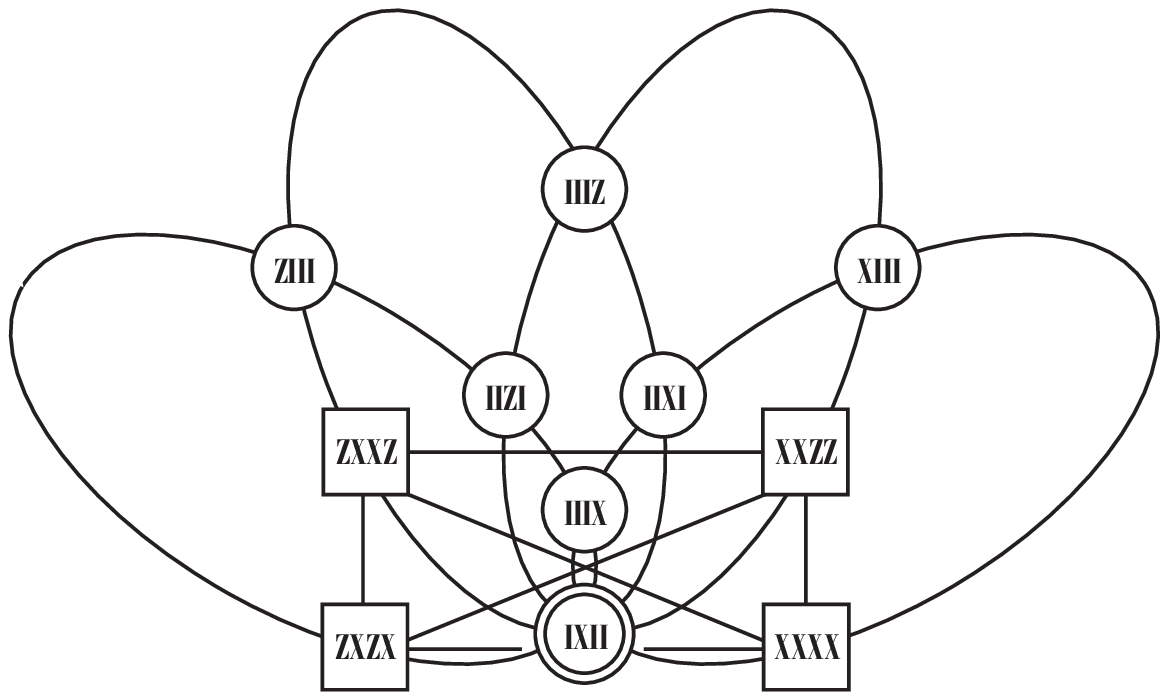}\includegraphics[width=7.3cm,clip=]{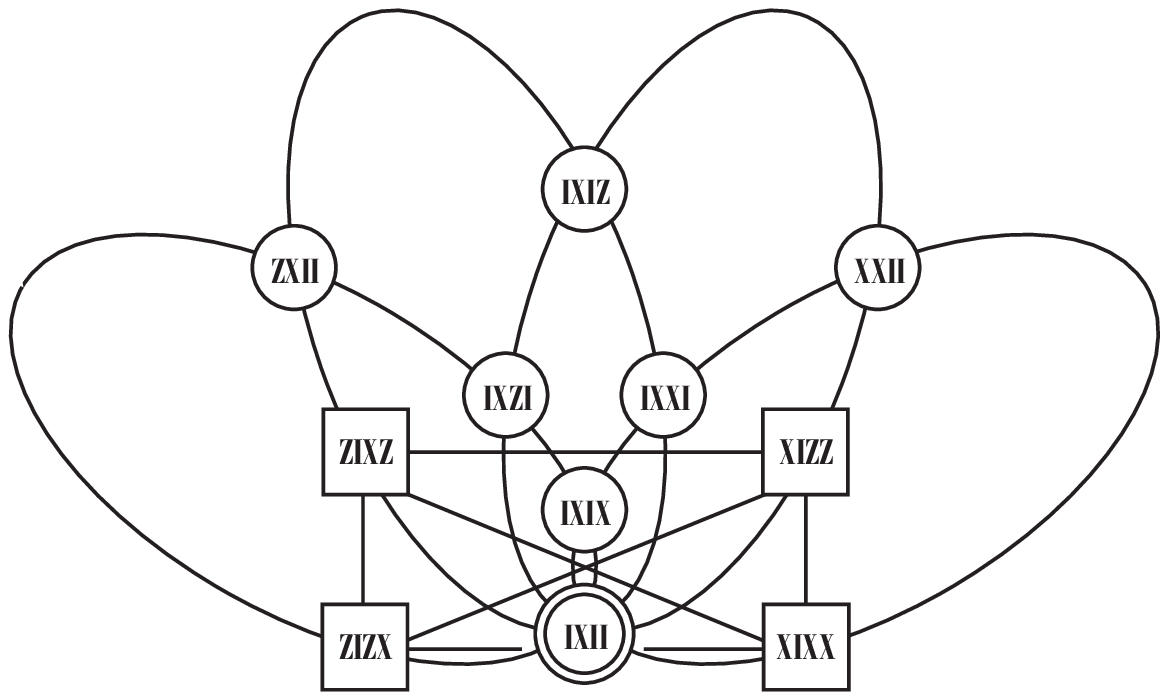}}

\vspace*{-5.0cm} \caption{{\it Left}: --- An illustration of the finite geometry behind the ``magic rectangle" of Harvey and Chryssanthacopoulos.
The four elliptic quadrics, generated by the first four sets, are represented by ellipses, whereas the affine plane of order two, underpinning $S_5$, is drawn as a quadrangle with two opposite pairs of points joined by line-segments as well (the most famous rendering of this plane). Note that, apart from the common point/observable $IXII$, the quadrics share pairwise one more point.
 {\it Right}:
--- A picture of the configuration that is, from the geometrical point of view, complementary to the previous one; the two configurations are, so to say, in perspective from the point/observable $IXII$.}
\end{figure}

In the final step, we first observe that the PG$(2,2)$ has a single point in common with each PG$(3,2)_i$. Next, a closer look at the sets of observables of the four PG$(3,2)$s themselves reveals that they pairwise share triples of operators whose product is always $IIII$; that is, every pair of our PG$(3,2)$s is on a common line. The six lines we get this way look explicitly as follows ($L_{ij} \equiv$ PG$(3,2)_i \cap$ PG$(3,2)_j$)
\begin{eqnarray*}
& L_{12} = \{IXII, ZIII, ZXII\}, \\
& L_{13} = \{IXII, IIZI, IXZI\}, \\
& L_{14} = \{IXII, IIIX, IXIX\}, \\
& L_{23} = \{IXII, IIIZ, IXIZ\}, \\
& L_{24} = \{IXII, IIXI, IXXI\},  \\
& L_{34} = \{IXII, XIII, XXII\},
\end{eqnarray*}
and are found to pass through the same point, $IXII$. A crucial observation at this place is that these six lines ``project" our four quadrics into another set of four quadrics on the same
point $IXII$ (the point of perspectivity, so to say). That is, each line associates a particular point of our H-C configuration with a unique off-configuration point; $ZIII$ with $ZXII$, $IIZI$ with $IXZI$, etc. We shall get in this manner a complementary set of four elliptic quadrics and, so, four complementary five-element sets of observables, namely:
\begin{eqnarray*}
S'_1 = \{IXII, ZXII, IXZI, IXIX, ZIZX\}, \\
S'_2 = \{IXII, ZXII, IXIZ, IXXI, ZIXZ\}, \\
S'_3 = \{IXII, IXZI, IXIZ, XXII, XIZZ\}, \\
S'_4 = \{IXII, IXIX, IXXI, XXII, XIXX\}.
\end{eqnarray*}
Moreover, if we also project from this ``exceptional" point $IXII$ the four points of our affine plane, we obtain a complementary affine plane and, so, a complementary four-element set of observables, namely:
\begin{eqnarray*}
S'_5 = \{ZIZX, ZIXZ, XIZZ, XIXX\}.
\end{eqnarray*}
All in all, we thus arrive at a {\it complementary}, or {\it twin} configuration depicted in Figure 1, {\it right}, which is ``magical" in the same way as the original one. To conclude this section, one mentions in passing that both $S_i$ and $S'_i$, $i$ running from 1 to 4, have the same ambient totally isotropic PG$(3,2)_i$ and that both $S_5$ and $S'_5$ have the same projective closure, the line $\{IIYY, YIIY, YIYI\}$.

\section{Conclusion}
We have found and briefly described a remarkable finite-geometrical representation of the Harvey-Chryssanthacopoulos ``magic rectangle" of eleven observables \cite{hch} providing a proof of the BKS theorem in the Hilbert space of four qubits.  In striking  analogy to similar ``magic" configurations employing sets of two- and three-qubit observables, also in the present case we find  {\it prominent} finite geometries  --- an affine plane of order two and an elliptic quadric of the binary projective space of three dimensions --- behind the scene. Even more importantly, this finite-geometrical approach enabled us to reveal that such H-C four-qubit configurations always come in {\it complementary pairs} (see Figure 1). Our findings thus give another significant piece of support to the importance of finite geometries and their combinatorics for the field of quantum information theory.

\section*{Acknowledgements}
This work was partially supported by the VEGA grant agency project 2/0098/10. We are extremely grateful to our friend Petr Pracna for an electronic version of the figure.

\end{document}